# High Energy Neutrino Astronomy: Status and Perspectives

Christian Spiering

*DESY, Platanenallee 6, D-15738 Zeuthen, Germany*

**Abstract.** The year 2008 has witnessed remarkable steps in developing high energy neutrino telescopes. IceCube at the South Pole has been deployed with 40 of its planned 80 strings and reached half a cubic kilometer instrumented volume, in the Mediterranean Sea the "first-stage" neutrino telescope ANTARES has been completed and takes data with 12 strings. The next years will be key years for opening the neutrino window to the high energy universe. IceCube is presently entering a region with realistic discovery potential. Early discoveries (or non-discoveries) with IceCube will strongly influence the design and the estimated discovery chances of the Northern equivalent KM3NeT. Following theoretical estimates, cubic kilometer telescopes may just scratch the regions of discovery. Therefore detectors presently planned should reach sensitivities substantially beyond those of IceCube.

## INTRODUCTON

High-energy neutrinos must be emitted as a by-product of high-energy collisions of charged cosmic rays with matter. Since they can escape much denser celestial bodies than light, they can be tracers of processes which stay hidden to traditional astronomy. Different to gamma rays, neutrinos provide incontrovertible evidence for hadronic acceleration. But at the same time their extremely low reaction probability makes their detection extraordinarily difficult.

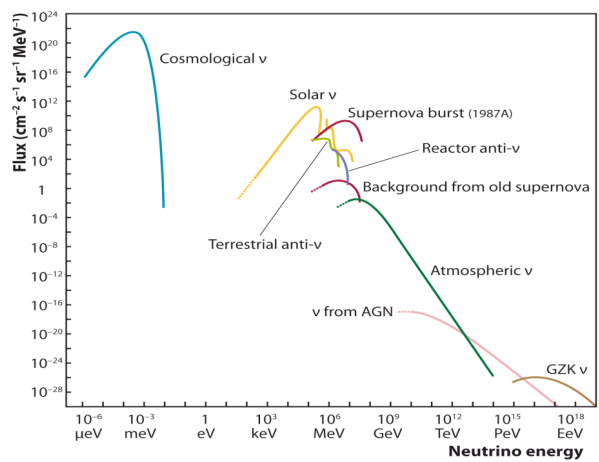

**Figure 1:** Spectra of natural and reactor neutrinos

Figure 1 shows a compilation of the spectra of dominant natural and artificial neutrino fluxes. Solar neutrinos, burst neutrinos from SN1987A, reactor neutrinos, terrestrial neutrinos and atmospheric neutrinos have been already detected. Another guaranteed – although not yet detected – flux is that of neutrinos generated in collisions of ultra-energetic protons with the 3K cosmic microwave background [1]. These neutrinos (marked GZK) as well as neutrinos from active galactic nuclei (marked AGN) or from other extraterrestrial sources will likely be detected by neutrino telescopes in the next decade. However, predictions for their fluxes are uncertain by orders of magnitude [2-7]. No practicable idea exists how to detect 1.9 K cosmological neutrinos.

Neutrino astrophysics is the central motivation to build neutrino telescopes on the cubic kilometer scale. Other topics include the indirect search for dark matter by searching for neutrinos produced in WIMP annihilation in the Sun or in the center of the Earth. Large neutrino telescopes may also open a new perspective for oscillation physics with atmospheric neutrinos. This includes "standard oscillations" as well the hypothetical effects of "non-standard oscillations" due to quantum de-coherence or violation of Lorenz invariance. They can also search for exotic particles like magnetic monopoles, super-symmetric Q-balls or nuclearites. Last but not least, the devices are also yielding interesting information on environmental effects – be it on deep natural water or Antarctic ice.

In accordance with the topic of this conference, this article focuses on the astrophysical aspect.

## SIGNAL PREDICTIONS

The neutrino flux recorded by a neutrino telescope consists of the following contributions:

a) atmospheric neutrinos which are generated in cosmic ray air showers in the Earth's atmosphere,
b) diffuse fluxes of extraterrestrial neutrinos,
c) point-like fluxes of extraterrestrial neutrinos.

Atmospheric neutrinos constitute a diffuse background to fluxes of extraterrestrial neutrinos. Compared to extraterrestrial sources, their energy spectrum is rather steep and follows an $E^{-3.7}$ behavior. Point-like sources can be identified by local excesses on top of the background of atmospheric neutrinos. Generic cosmic acceleration processes lead to an $E^{-2}$ spectrum, i.e. much harder than that of atmospheric neutrinos. Specific models result in more complicated energy dependencies, typically with cut-offs at lower and higher energies – the one due to threshold effects in generating neutrinos, the other due to cosmic accelerators running out of power. Point source searches may make use of the expected energy dependence. They may cut off or suppress the lower energies, thereby strongly reducing the background of atmospheric neutrinos, with a still acceptable loss of signal events. Point source searches may also make use of variations in the flux from certain cosmic objects and focus the search to those time windows where the expected neutrino flux, and correspondingly the signal-to-noise ratio, is high. For *diffuse searches*, only one of the three criteria (direction, spectrum, variability) remains: the energy spectrum. Therefore, in order to suppress the background from atmospheric neutrinos, diffuse searches have to apply much tighter energy cuts than point-source searches. Naturally, the fluxes from such extreme extragalactic objects like AGN or GRB (Gamma Ray Bursts) are expected to extend towards much higher energies than those from galactic accelerators. Therefore a cut at very high energies would still let pass a significant part of the extragalactic signal itself.

Whereas galactic sources may mostly reveal themselves as point sources, most of the extragalactic flux will appear as a diffuse, isotropic flux, with the exception of rather close, bright sources. Given existing limits on diffuse fluxes, the latter fact has important implications for the flux expectation from point sources [5] – see below.

High energy neutrino production is related to the production of charged cosmic rays and gamma rays. From the observed fluxes of cosmic rays and gamma rays, estimates or upper bounds on the flux of neutrinos can be derived. Since high-energy gamma rays may well emerge from inverse Compton scattering, their observation is not a proof that the source accelerates hadrons rather than only electrons. A certain test can be provided by detailed information on the MeV-GeV part of the gamma spectrum and by information on its high energy cut-off. For galactic sources, the morphology of gamma ray emission can be studied and provides additional information. In this context, we note the observation of the Supernova Remnant (SNR) RX J1713.7-3946 by the H.E.S.S. telescope [8]. The image shows an increase of the gamma flux from the direction of known molecular clouds. The effect may be related to protons accelerated in the SNR and then interacting with the clouds. The decay of the produced $\pi^0$s would contribute to the gamma ray emission. A similar extended gamma ray source which traces the density of molecular clouds has been identified near the galactic center [9]. The ultimate, water-tight demonstration of hadronic acceleration would be, however, the observation of neutrinos.

### *Galactic Sources*

The preferred candidates for galactic sources are Supernova Remnants (SNR), pulsar wind nebulae and compact binary systems. A series of papers has appeared over the last years [5, 10-15], all with the unanimous conclusion that cubic kilometer detectors will just "scrape" the detection region. These calculations assume that most of the observed gamma rays stem from $\pi^0$ decays and then recalculate the neutrino flux expected from $\pi^{+/-}$ decays. From the observed high-energy cut-offs of several gamma spectra at ~ 100 TeV follows a somewhat lower high-energy cut-off for the neutrino spectra. A selection of high energy neutrinos helps to suppress the background from atmospheric neutrinos, but the cut should not be higher than ~5-10 TeV since otherwise the signal is lost.

In [13], the neutrino flux from the SNR RX J1713.7-3946 is calculated, see Fig.2. Assuming five years of data taking with a cubic kilometer detector, the number of events turns out to be between 7 and 14, on a background of 21 atmospheric neutrino events in a 1.3° search bin. This assumes a threshold at 1 TeV. A 5-TeV threshold results in 2.6-6.7 signal events on a background of 8.2 atmospheric neutrino events. Cutting at higher energies will eliminate not only the atmospheric background but also the signal, cutting at lower energies will significantly worsen the signal-to-background ratio. These estimates suggest that the positive effect of a detector threshold much below one TeV will be small, at least for steady sources.

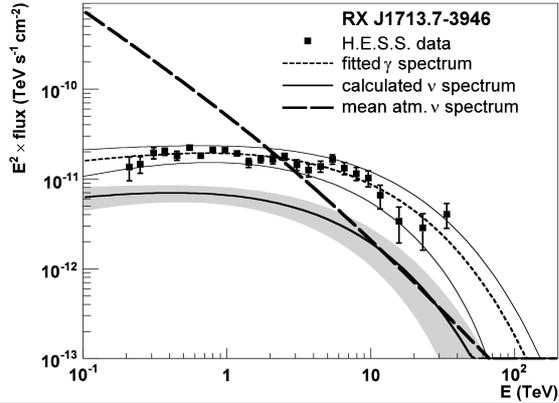

**Figure 2:** Measured gamma flux and estimated neutrino flux from RX J1713.7-3946 [13]. Courtesy A. Kappes.

The position of the "knee" in the cosmic ray spectrum suggests that some sources accelerate cosmic rays to energies of several PeV. These "Pevatrons" may produce, via the decay of $\pi^0$s, gamma rays whose spectrum extends to several hundred TeV. Recently, candidates for such Pevatrons have been identified by the MILAGRO collaboration [16,17]. The gamma ray spectrum of the strongest of these sources, MGROJ1908+06, is consistent with an $E^{-2}$ behavior between 500 GeV and 40 TeV and shows no evidence for a cut-off. In [15], the associated neutrino flux from the five identified MILAGRO source candidates has been calculated and the events rates in the IceCube neutrino telescope estimated. In a simulated neutrino sky-map for 5 years data taking, two of the sources are merely visible "by eye", applying an energy threshold of 40 TeV. Stacking all six sources, a Poisson probability of the excess smaller than $10^{-3}$ could be obtained within 5 years if a low-energy cut anywhere between 10 and 100 TeV is applied. The simulation assumes a gamma ray cut-off of the sources at 300 TeV.

Summarizing, the present estimates suggest the sensitivity of a cubic kilometer telescope being "tantalizingly (and frustratingly) close" [5] to the expectations for the brightest observed galactic TeV gamma sources.

*Extragalactic Sources*

There is much less experimental guidance to predict neutrino fluxes from extraterrestrial sources than for galactic sources. The predictions for individual sources are sometimes considered as being "still at the level of reading tea leaves" [18]. Predictions for the integrated flux from all extragalactic sources are based on the observed fluxes of gamma or X-rays or of charged cosmic rays above $10^{18}$ eV. Early normalizations as e.g. [19] assumed neutrinos would be produced in the cores of AGN, accompanied by X-ray emission, and that most of the observed X-ray background was non-thermal radiation from a superposition of the fluxes from unresolved AGN. This model as well as others obviously violated upper bounds derived from the observed cosmic ray spectrum. Subsequent observations of these AGN showed that most of the X-ray emission is thermal and therefore cannot be directly related to the production of relativistic particles. Relying on MeV gamma ray rather than X-ray observations, the authors scaled down the original prediction by an order of magnitude ([20], *StSa* in Fig. 3).

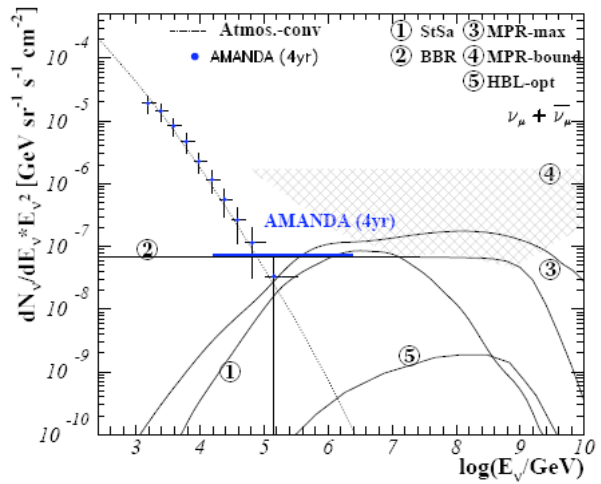

**Figure 3:** AGN neutrinos flux models (taken from [22]): (1) normalization to MeV gamma rays [20], (2) normalization to radio emission from FR galaxies [24], (3) maximum contribution from EGRET sources, (4) upper bounds [21] for from sources optically thick for neutrons (upper straight bound) and optically thin for neutrons (lower curved bound of shaded area). The Waxman-Bahcall bound would be a horizontal line at $10^{-8}$ (max.$5 \cdot 10^{-8}$). (5) prediction for high-peaked BL-Lacs within the proton blazar model [25]. The prediction for atmospheric neutrinos [26] is shown as a dotted line. The data points are for four years of AMANDA data [27], the limit on a possible $E^{-2}$ contribution from extraterrestrial neutrinos to the measured AMANDA data is given by the horizontal line [28]. Courtesy J. Becker.

Actually two bounds, both derived from charged cosmic ray fluxes, have been frequently used as benchmark bounds. The first ("Waxmann-Bahcall bound") has been derived in [3] and is normalized to the cosmic ray flux at $\sim 10^{19}$ eV. Assuming a generic $E^{-2}$ spectrum for all extragalactic sources, the authors obtain an upper limit of $E_\nu^2 \cdot dN/dE_\nu = 1\sim5 \times 10^{-8}$ GeV cm$^{-2}$ s$^{-1}$ sr$^{-1}$, with the uncertainty given by different cosmic evolution models. In a cubic kilometer detector, this would lead to 100-500 events per year.

Contrary to Waxmann/Bahcall, Mannheim, Protheroe and Rachen [21] assumed that a significant part of the observed cosmic ray spectrum below $10^{19}$ eV could be due to extra-galactic rather than only galactic sources. Interpreting the cosmic ray spectrum below $10^{19}$ eV as a convolution of spectra from many extragalactic source classes, each with a different cut-off, the neutrino bound considerably weakens and is about $E_\nu^2 \cdot dN/dE_\nu \sim 5 \times 10^{-7}$ GeV cm$^{-2}$ s$^{-1}$ at a few $10^{14}$ eV. Figure 3 (taken from [22]) shows a compilation of different model predictions and bounds.

The most exiting discovery would be the detection of point sources rather than just a clear high-energy excess in diffuse fluxes. However, experimental limits for diffuse fluxes are setting bounds for expected point-source fluxes [5]. The argument is the following: contributions to the diffuse flux will come from all the observable universe, up to a distance $c/H_0$, whereas point sources, with several events per source, will be visible only up to a limited distance of a few hundred Mpc, assuming reasonable maximum luminosities per source. For a homogeneous distribution of extra-galactic sources, one therefore can derive a limit on the number of observable point sources. In [5] the following assumptions are made: a) a homogeneous source density in a Euclidian universe, b) a source luminosity $L_{source}$ "typical" (and similar) for all sources, c) an $E^{-2}$ behavior of sources. Then, assuming an experimental limit $K_{diffuse}$ on the diffuse flux and a sensitivity $C_{pt}$ to point sources, the expected number of resolvable extragalactic point sources, $N_s$ is

$$N_s \sim \frac{K_{diffuse} \cdot \sqrt{L_{source}}}{C_{pt}^{2/3}}$$

With the present diffuse limit from AMANDA and the expected point source sensitivity of IceCube, one obtains $N_s \sim 1\text{-}10$ [23]. This means that, with the given assumptions, a cubic kilometer detector would have a fair – but not overwhelmingly large – chance to detect extragalactic point sources! Note however, that even if IceCube would push $C_{point}$ twenty times below that of AMANDA, a few individual, very close sources could circumvent the homogeneity assumption and be well observable. Also, point sources with cut-offs below a few hundred TeV would not be covered by the argument above since, in order to obtain the best sensitivity for *diffuse* fluxes, IceCube will place energy cuts at about 100 TeV [36].

## DEVICES AND RESULTS

Neutrino detectors underground have opened the neutrino window at low energies [29]. In order to detect the low fluxes from the suspected distant sources of *high-energy* neutrinos, detectors of cubic kilometer volume or more are required. They cannot be arranged underground but only in deep oceans, lakes or glacial ice where available space is no issue. In these transparent media, the Cherenkov light emitted by charged secondary particles from neutrino interactions is registered by light sensors spread over a large volume. This principle was first proposed in 1960 [30]. In the mean time, first-generation detectors are operated in the Siberian Lake Baikal (NT200) [31], in Antarctic ice (AMANDA) [32, 33] and since recently in the Mediterranean Sea (ANTARES) [34,35]. Moreover, half of a second-generation telescope of cubic kilometer size, IceCube [36], has been installed and takes data.

Figure 4 sketches the two basic detection modes of underwater/ice neutrino telescopes. Charged current muon neutrino interactions produce a muon track (left). Apart from elongated muon tracks, a neutrino interaction may also lead to cascades (right). Charged current electron and tau neutrino interactions transfer most of the neutrino energy to electromagnetic cascades initiated by the final state electron or tau, hadronic cascades are the product of the target hadron disintegration or hadron tau decays. In most models, neutrinos are produced in a ratio $\nu_e:\nu_\mu:\nu_\tau \approx 1:2:0$. Over cosmic distances, oscillations turn this ratio to $\nu_e:\nu_\mu:\nu_\tau \approx 1:1:1$, which means that about 2/3 of the charged current interactions appear as cascades.

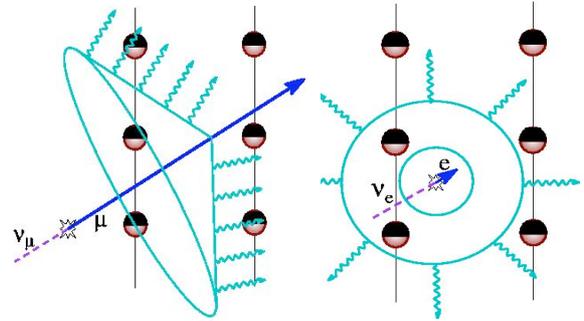

**Figure 4:** Detection of muon tracks (left) and of cascades (right) in underwater/ice detectors.

The classical operation is recording upward traveling muons, since only neutrinos can cross the Earth. A neutrino telescope must be arranged at > 1 km depth in order to suppress the background from misreconstructed downward moving muons. After suppressing this background, only one unavoidable background to extraterrestrial neutrinos remains: neutrinos generated by cosmic ray interactions in the Earth's atmosphere ("atmospheric neutrinos"). This background cannot be removed by going deeper. On the other hand, it provides a standard candle and a reliable proof of principle.

Optical detectors are tailored to TeV and PeV energies. Towards even higher energies (> 100 PeV), novel detectors aim detecting the coherent Cherenkov radio signals (ice, salt) or acoustic signals (water, ice, salt) from neutrino-induced showers. Air shower detectors search for showers with a "neutrino signature". The very highest energies are covered by balloon-borne detectors recording radio emission in terrestrial ice masses, by ground-based radio antennas sensitive to radio emission in the moon crust, or by satellite detectors searching for fluorescence light from neutrino-induced air showers. This article focuses on optical detectors in water and ice.

## First Generation Telescopes

### *The Lake Baikal Neutrino Telescope*

The Baikal Neutrino Telescope is installed in Lake Baikal at a depth of about 1.1 km. The BAIKAL collaboration was not only the first to deploy, in 1993, three strings (as necessary for full spatial reconstruction) [37], but also reported the first atmospheric neutrino detected underwater.

The central part of the Baikal configuration is NT200, an array of 192 optical modules (OMs) which was completed in April 1998 and takes data since then. The geometrical dimensions of the configuration are 72 m (height) and 43 m (diameter). Due to the small lever arm, the angular resolution of NT200 for muon tracks is only 4°. The small spacing implies a muon energy threshold as low as ~15 GeV. The total number of upward muon events collected over 5 years is ~400. Still, NT200 could compete with the much larger AMANDA for a while by searching for high energy cascades *below* NT200, surveying a volume about ten times as large as NT200 itself. In order to improve pattern recognition in this volume, it was fenced in 2005-07 with 3 sparsely instrumented outer strings. This configuration was named NT200+ [38].

### *AMANDA*

Rather than water, AMANDA (Antarctic Muon And Neutrino Detection Array) uses the 3 km thick ice layer at the South Pole as target and detection medium. The array was completed in January 2000 and comprises 19 strings with a total of 677 OM, most of them at depth between 1500 and 2000 m [32, 33].

The angular resolution of AMANDA for muon tracks is 2°-2.5°, with an energy threshold of ~50 GeV. Although better than for Lake Baikal (4°), it is much worse than for ANTARES (<0.5°, see below). This is the result of the strong light scattering in ice which deteriorates the original information contained in the Cherenkov cone. The effect is even worse for cascades, where the angular resolution achieved with present algorithms is only ~25° (compared to 5°-10° in water). The advantages of ice compared to water are the larger absorption length and the small PM noise rate, about 1 kHz in an 8-inch PM, compared to 20-40 kHz due to $K^{40}$ decays and bio-luminescence in lakes and oceans. This makes hit cleaning procedures much easier than in water.

### *ANTARES*

ANTARES [40] is operating close to Toulon in the Mediterranean Sea, at a depth of nearly 2500 m. It started data taking with five strings in 2007 and has been completed in 2008. It now consists of 12 strings, separated horizontally by 60-75 m. Each string is instrumented, over 350 m length, with 25 "storeys". A storey is equipped with three 10-inch PMs housed in 13-inch glass spheres. The PMs are oriented at 45° with respect to the vertical.

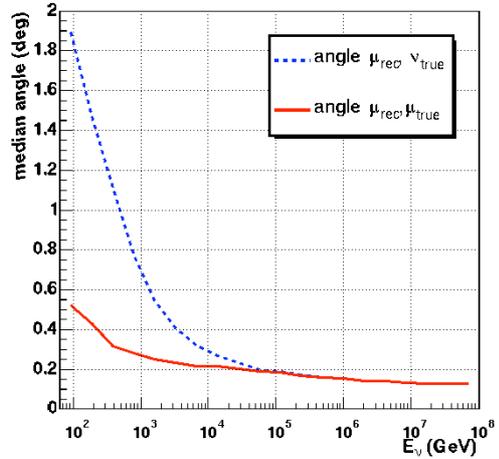

**Figure 5:** ANTARES angular resolution as a function of the neutrino energy [41]. Courtesy T. Montaruli.

The Monte Carlo angular resolution for muons is 0.3 degree at 1 TeV. In practice, the limited knowledge of absolute position and of varying detector parameters will make the further improvement towards higher energies a challenging task. Naturally, the angular resolution for cascades is worse than for muons. Simple reconstruction algorithms give 10° median mismatch angle above 5 TeV, however, with proper quality cuts, values below 5° can be achieved, with 20-40% passing rates for signals [42].

*Physics Results From First-Generation Telescopes*

Naturally, the wealth of results has been obtained by AMANDA and NT200. A full list of references to the results listed below can be found, for instance, in [43] for the Baikal experiment, and in [39,44,45] for AMANDA.

- **Atmospheric neutrinos:** All three experiments have shown that they can reliably separate upward muons from downward muons and that Monte-Carlo simulations agree with experimental data both below the horizon (with atmospheric neutrinos providing the standard candle) and – typically somewhat worse – above the horizon. As an example we show in Fig.6 the muon angular distribution obtained from ANTARES [46].

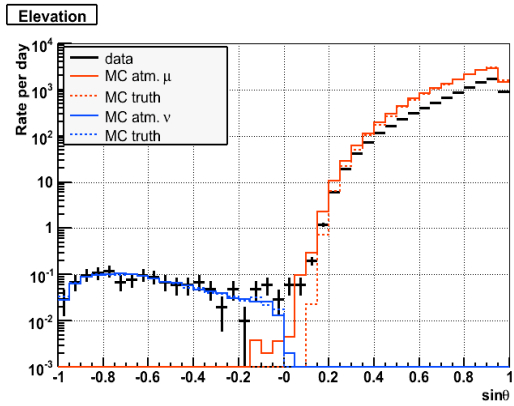

**Figure 6:** Distribution of sin(zenith) for data taken with the first 5 strings of ANTARES compared to MC. Solid line for reconstructed, dashed line for true muons [46]. Courtesy J. Brunner.

With 6595 neutrinos collected in seven years, AMANDA measured the spectrum of atmospheric neutrinos up to about 100 TeV (two orders of magnitude beyond energies covered by underground experiments, see Fig. 3) and is going to establish limits on non-standard oscillations [39].

- **Point sources:** The upper limit on the flux from extraterrestrial point sources of muon neutrinos as measured with AMANDA over seven years [47] is $5 \cdot 10^{-11}$ $E_\nu^{-2}$ TeV$^{-1}$ cm$^{-2}$ s$^{-1}$ sr$^{-1}$ (averaged over the Northern hemisphere), more than an order of magnitude below the limits measured with underground detectors on the Southern hemisphere. For extremely high-energy point sources ($10^6$-$10^8$ GeV) the analysis even extends into the Southern hemisphere, with a sensitivity of $(3-10) \cdot 10^{-10}$ $E_\nu^{-2}$ TeV$^{-1}$ cm$^{-2}$ s$^{-1}$ close to the horizon [39, 45].

Figure 7 shows the sky map derived from seven years AMANDA data taking and 6595 events [47]. All observed spots are compatible with background fluctuations, resulting in the limit given above.

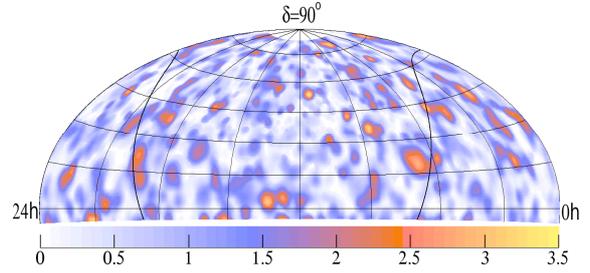

**Figure 7:** Seven-year AMANDA sky map. Colours indicate significances *before* correction for trial factors. The highest pre-trial significance of 3.38σ has a 95% chance to appear in simulated sky-maps.

- **Diffuse extraterrestrial fluxes:** Figure 8 is a compilation of existing limits on diffuse fluxes. The AMANDA upper limits on the extraterrestrial diffuse flux of energetic neutrinos (normalized to the flux from one flavor) are $7.4 \cdot 10^{-8}$ $E_\nu^{-2}$ GeV$^{-1}$ cm$^{-2}$ s$^{-1}$ sr$^{-1}$ ($10^4$-$10^6$ GeV muons), $2.9 \cdot 10^{-7}$ $E_\nu^{-2}$ GeV$^{-1}$ cm$^{-2}$ s$^{-1}$ sr$^{-1}$ ($10^4$-$10^7$ GeV cascades) and $3.3 \cdot 10^{-7}$ $E_\nu^{-2}$ GeV$^{-1}$ cm$^{-2}$ s$^{-1}$ sr$^{-1}$ ($10^5$-$10^9$ GeV muons). The Baikal limit is $2.7 \cdot 10^{-7}$ $E_\nu^{-2}$ GeV$^{-1}$ cm$^{-2}$ s$^{-1}$ sr$^{-1}$ ($10^5$-$10^8$ GeV). These limits exclude several models on neutrino production in AGN (see [52] for a discussion of astrophysical implications).

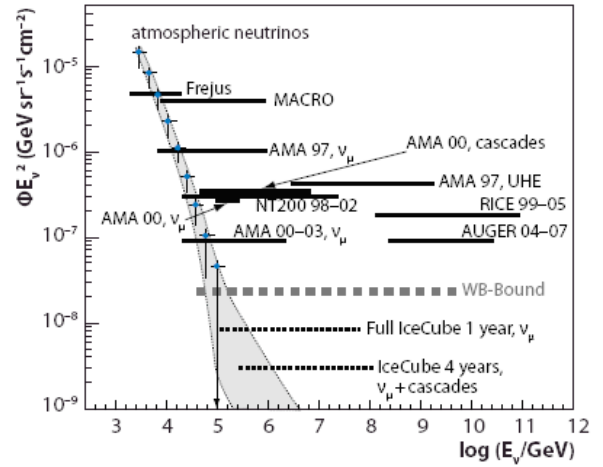

**Figure 8:** Upper limits on the diffuse flux of extraterrestrial neutrinos with an $E^{-2}$ spectrum. All all-flavor limits have been divided by 3, assuming a 1:1:1 ratio of neutrino flavors at Earth. The grey area indicates uncertainties of the atmospheric neutrino flux. Modified from [48], where further explanations and references can be found.

Note that future diffuse searches are going to be challenged by the limited knowledge on the contribution of prompt atmospheric neutrinos from charm decay. Combination of muon and cascade data may help to disentangle this contribution from that of extraterrestrial neutrinos [51].

- **Gamma Ray Bursts:** AMANDA data have been searched for coincidences between neutrinos and GRB reported by BATSE, IPN and SWIFT. No coincidences have been observed, and an upper limit of $6 \cdot 10^{-9} E_\nu^{-2} GeV^{-1} cm^{-2} s^{-1}$ on the flux of neutrinos coincident with GRB has been derived [49]. With this limit, AMANDA approaches the Waxman-Bahcall GRB model prediction [50] by less than a factor 2. IceCube can test this prediction within several months.

- **Indirect WIMP search:** The upper limits on the flux of muons produced by neutrinos from WIMP annihilation in the center of the Earth (AMANDA and NT200) and the Sun (only AMANDA) are about or already below best underground limits: $0.8 \cdot 10^3$ km$^{-3}$ year$^{-1}$ for WIMP masses above a few hundred GeV. Assuming standard SUSY parameters, the solar flux limits are close within a factor 3 to the limits by direct searches.

- **Relativistic monopoles:** The upper limits on the flux of relativistic monopoles $dF_M = 2.8 \ (4.6.) \cdot 10^{-17}$ GeV$^{-1}$ cm$^{-2}$ s$^{-1}$ sr$^{-1}$ for AMANDA (NT200) are 20 times below the Parker limit and 3 times below best underground limits.

- **Slow monopoles:** The Baikal limit on the flux of slow monopoles catalyzing proton decay is $2 [50] \cdot 10^{-16}$ GeV$^{-1}$cm$^{-2}$s$^{-1}$sr$^{-1}$, assuming a catalysis cross section $10^{-28}$ [$10^{-30}$] cm$^2$ and a velocity $\sim 10^{-4} c$.

- **Supernova Burst Monitoring:** Due to the low ambient light noise, AMANDA is sensitive to MeV neutrinos for Supernova bursts – by detecting the feeble increase in PMT counting rates over a time interval over several seconds. AMANDA has been monitoring the Galaxy for supernova bursts over several years and provides input to the Supernova Early Warning System, SNEWS.

- **Cosmic Rays:** The South Pole Air Shower Array, SPASE, operated together with AMANDA, confirmed the increase in the mass composition of cosmic rays above the knee.

## Second Generation Telescopes

The construction of detectors on the cubic kilometre scale is ongoing, or being prepared, at three locations: at the South Pole (IceCube), in the Mediterranean Sea (KM3NeT) and in Lake Baikal (GVD).

*IceCube*

IceCube will consist of 4800 Digital Optical Modules (DOMs), attached to 80 strings at depths between 1450 and 2450 m [53]. A DOM consists of a pressure glass sphere housing a 10-inch diameter Hamamatsu photomultiplier plus associated electronics. The strings are arranged on a hexagonal grid with 125 m spacing, covering 1 km$^2$. IceCube is complemented by the air shower detector IceTop on the surface, made of 160 ice-filled tanks, two near the top of each string. One string has been deployed in the season 2004/05, eight in 2005/06, 13 in 2006/07 and 18 in 2007/08. Forty strings, i.e. 50% of IceCube, are now deployed and take data. This will allow doing physics at the cubic kilometre scale starting in 2009. With 80 tanks on the surface, IceTop is also 50% completed.

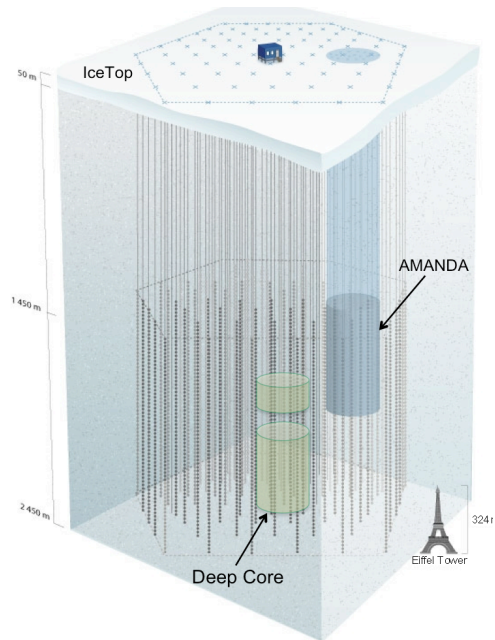

**Figure 9:** The IceCube Observatory, consisting of IceCube and IceTop plus a low-energy sub-detector – at present AMANDA, in future DeepCore.

AMANDA is meanwhile integrated into IceCube and serves as a dense core with low-energy threshold. It is foreseen to be replaced by DeepCore, a 6-string, 360-PMT array with small spacing located at the bottom centre of IceCube, below 2100 m, in the best ice (see Fig. 9). AMANDA will be decommissioned in 2009 due to the snow accumulation around the AMANDA counting house, its high power consumption and its over-proportional maintenance effort when compared to IceCube. Compared to AMANDA, the DeepCore performance would be significantly improved, since the rest of IceCube serves as efficient veto and since it is located in better ice. DeepCore would allow observation of the Southern sky, extension of the indirect dark matter search to lower WIMP masses, slightly improved point source sensitivity for steep source spectra and, possibly, oscillation studies in a hitherto unexplored energy regime [54].

The IceCube hardware is working extremely well [55,56]. The rate of registered neutrinos increased from 1.5/day (9 strings) to 20/day (22 strings) and is projected to be 200/day for the full detector. The 2007 data (22 strings) are presently in the final stage of analysis. Overviews on results from 2006 data and preliminary 2007 data can be found in [54, 56]. We list some of them:

- The point-source sensitivity of one year of the 22-string configuration (~1/4 IceCube) exceeds that of seven years AMANDA by a factor 1.5-3, depending on the analysis [63]. The 2007 sky-map is compatible with background. Figure 10 includes an estimate of the corresponding average limit as well as extrapolations to the exclusion sensitivities expected in a few years from now. Note that $5\sigma$ discovery sensitivities are typically a factor 2.5 worse than average 90% CL. limits. A central message of Fig. 10 is that in 2012 the sensitivities to point sources will have improved by three orders of magnitude when compared to the situation ten years ago.

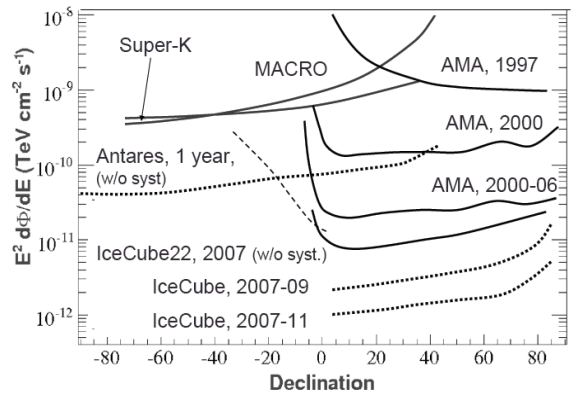

**Figure 10:** Existing average 90% C.L. upper limits (full lines) and expected sensitivities to $E^{-2}$ sources vs. declination. Limits for individual sources published by Super-K [57] and MACRO [58] have been interpolated to better guide the eye. The ANTARES curve has been taken from [59]. The IceCube 22-string average limit sensitivity is preliminary and does not include systematic uncertainties. The dashed line indicates the extension of the IceCube-22 sensitivity towards the Southern hemisphere when optimizing the analysis to high energies.

- Among transient sources, Gamma Ray Bursts (GRB) are the most prominent. During the recent extremely bright burst at March 19 (GRB080318B) IceCube was running in a 9-string maintenance mode. A fireball model with a Lorentz boost of 300 predicts 0.1 events for this configuration – an analysis is underway.
- A search for high-energy neutrinos from the Sun was performed with IceCube-22, using 4 months of data when the Sun was below horizon. No excess over atmospheric neutrinos was found, resulting in a limit of about 350 muons with $E > 1$ GeV per $km^2$ and year. This improves the previous limit from one year AMANDA by about a factor 2 and approaches the expected 5-year IceCube limit within a factor of about 4.
- IceTop is being used to measure the cosmic ray energy spectrum and its composition. The measured spectrum (year 2006) agrees well with the spectral index 3.05 found by other experiments. An analysis of the angular dependence favors a mixed composition. The full power of these investigations will be provided by measuring coincident events in IceTop and IceCube [60, 61].
- Although the IceTop array has a threshold of about 300 TeV, individual tanks are sensitive to single particles from cosmic air showers of much lower energies which can increase the counting rates in the tanks. Therefore the tanks are sensitive to particles from solar flares. Actually, on Dec. 13, 2006, a solar flare observed by monitoring stations all around the world led to a 1% increase of IceTop counting rates over half an hour. Since the rate depends on individual tank thresholds, even the energy spectrum could be inferred [62].

A large number of analyses is presently underway, starting with methodical aspects (like using the moon shadow for cosmic rays to determine absolute pointing and angular resolution of the 40-string detector) and ending with new results on steady and transient point sources, diffuse extraterrestrial neutrino fluxes, atmospheric neutrinos, dark matter and exotic particle searches. Several multi-messenger campaigns are under preparation and will be discussed below. The year 2009 promises a wealth of new data and analyses peering into new sensitivity regions.

*KM3NeT*

To complement IceCube, the three Mediterranean neutrino projects ANTARES, NEMO and NESTOR have joined their forces. Efforts in preparation of a neutrino telescope at the cubic kilometer scale, KM3NeT [64] are related to three locations.

- a site close to Pylos at the Peloponnesus, with available depths ranging from 3.7-5.2 km for distances to shore of 15 to 48 km (project NESTOR [65,66]),
- a site close to Sicily, at a depth of 3.5 km and 70 km distance to shore (project NEMO [67, 68]),
- a site close to Toulon, at a depth of 2.5 km and 40 km distance to shore (project ANTARES [34, 40, 69] ).

All of these sites are considered to be possible locations for a future cubic kilometer array. All sites have physics and infrastructural pros and cons. For instance, large depth is a challenge for long-term ocean technology and bears corresponding risks, but has convincing physics advantages: less background from punch-through muons from above and less bio-luminescence. Different distances to shore result in different optimum solutions for data transmission. Deployment procedures depend on the available ships, possibly also remote underwater vehicles, where the latter are available only down to certain depths. Last but not least, for a final site selection regional funding sources and political strategies may play an equally strong role as scientific arguments.

Whereas ANTARES is presently operated as a first-stage neutrino telescope, NESTOR and NEMO have operated single components or sub-detectors in an R&D mode. It is obvious that a simple scale-up of the three pilot projects to cubic kilometer scale is neither technically feasible nor financially possible. The present Design Study for KM3NeT aims to deliver the specifications for a detector which yields the best physics for a given budget. A Technical Design Report is planned for late 2009. Political decisions towards endorsing and funding KM3NeT are addressed in a "Preparatory Phase" project which recently has been funded by the European Union.

The declared minimal performance objective of KM3NeT is an effective volume of 1 km$^3$ with an angular resolution of 0.1° for muons with energies above 10 TeV. The detector has to be sensitive to all neutrino flavors, and the lower energy threshold should be at a few hundred GeV [70-72]. The envisaged overall costs are estimated as 220-250 M€. I will discuss these benchmarks in the final section, together with the corresponding recommendations of the ApPEC roadmap committee.

Detailed simulations have been performed for detectors of cubic kilometer size, using up to 10 000 optical modules. The effective area turns out to be about a factor 2 higher than that of IceCube, mainly due to the larger total photocathode area. Together with a better angular resolution, this would transform to a 2-3 times better sensitivity to point sources. Whether this could be called a substantial improvement of sensitivity compared to IceCube depends on whether IceCube sees or does not see sources. In case IceCube would fail to identify sources, and given the uncertainty of flux predictions, a "substantial" improvement should be larger than only a factor 2-3. This may require to sacrifice low-energy performance and to take a consequent decision for best high-energy capability.

A number of technologically interesting solutions have been developed within the three separate pilot projects and the common design study, for instance:

- **Towers vs. strings:** ANTARES uses a conventional string structure. It may meet difficulties if a high density of photocathode area is envisaged and separate strings come too close to each other. The NESTOR design uses towers composed of hexagonal floors. A floor with twelve 15-inch PMs has been deployed in 2003, but the shore cable failed after a few weeks of data taking. The basic units of NEMO are towers composed by a sequence of floors. The floors consist of rigid horizontal structures, 15 m long, each equipped with four 10-inch PMs. The floors are tilted against each other and form a three-dimensional structure. The use of "compacted" detector units which unfold during deployment, each housed in a container, several of which are being interconnected prior to deployment, would strongly simplify and speed up the deployment.
- A special deployment platform ("Delta Beriniki") has been developed within the NESTOR project.
- **Photo-detectors:** apart from configuring optical modules with one 10-inch PMT per module in various ways, a single optical module may house many 3-inch PMTs [73] achieving a higher photo-cathode area per module. Other options being investigated are multi-anode PMTs or hybrid photo-detectors (X-HPDs). HPDs obey better solid angle coverage than large standard PMTs, resulting in a better light collection and larger maximum spacing between optical modules.
- Various innovative methods for data reduction, fast data transmission to shore and data acquisition at shore are being tested.
- In order to avoid the expensive use of titanium for the structures and pressure-resistant components, the NEMO collaboration has developed a junction box which consists of steel vessels housed in an oil-filled fiber-glass container. This avoids direct contact between steel and the aggressive sea water and is cheaper than using titanium.

The envisaged full start of KM3NeT construction is 2011, with a 3 year deployment period. I note that the 3-year period seems extremely tight, given the deployment and budget challenges.

*GVD*

The Baikal Collaboration plans the stepwise installation of a kilometer-scale array in Lake Baikal, the Giant Volume Detector, GVD [74]. It would consist of 90-100 sparsely instrumented strings. Only 12-16 OMs would be arranged over 350 m string length (the comparatively shallow depth does not allow larger vertical dimensions). Each four strings would form a triangular cell of 200 m side length, with the fourth string in the centre. The overall horizontal dimensions would be close to two square kilometers, and the geometrical volume ~ 0.7 km³. Given the small absorption length and the large spacing, the threshold for muons is as high as ~5 TeV which coincides with the offline threshold which anyway has to be applied in order to get the best signal-to-noise (extraterrestrial versus atmospheric neutrinos) for the weakest detectable sources. The stepwise construction of GVD is planned to start around 2011 and to be completed within five years.

## MULTI-MESSENGER METHODS

The "classical" example for multi-messenger analyses is the search for neutrino events coinciding in direction and time with Gamma Ray Bursts reported from satellite experiments. Reduction of the search interval windows to short time windows around the GRB time considerably reduces the accumulated background from atmospheric neutrinos, allows for relaxed cuts in the analysis and enhances the resulting effective area of the neutrino telescope. Another example is the consolidation of a neutrino signal by a subsequent optical observation of a supernova, like for the supernova 1987A, or an alert to optical telescopes. Present multi-messenger approaches for neutrino telescopes go beyond these examples and resemble the successful multi-wavelength approach of classical astronomy [22, 75]. Three main rationales are behind multi-messenger methods:
- The selection of sources already known from electromagnetic waves reduces the trial factor penalty arising from observation of multiple sky bins.
- The significance of a neutrino observation can be enhanced by a coincident observation in X-rays, gamma-rays or from other electromagnetic wavelength bands.
- The combination of observations with different messengers provides more complete information on the cosmic source.

A strong motivation for multi-messenger approaches was given by the observation of two neutrino events from the direction of the Blazar 1ES1959+650, recorded by AMANDA in 2002 and coinciding with gamma ray flares recorded by Whipple and HEGRA [75]. The *a posteriori* determination of the statistical significance of the observation turned out to be problematic – the chance probability may have been on the percent level. One subsequent phenomenological paper claimed that the observation could be due to a real signal [76], another one demonstrated that this was practically excluded [77]. However, the main consequence of the observation was the idea to trigger gamma ray observations if a neutrino event from one of a few pre-selected source directions was observed. This method was christened Neutrino-Triggered Target of Opportunity (NToO). NToO operation not only may enhance the neutrino discovery chance but also increases the availability of simultaneous observations, complementing X-ray ToO triggers for gamma telescopes. A first technical NToO implementation between AMANDA and MAGIC was tested for a few pre-selected sources in a short run (Sept-Dec. 2006), showing the feasibility of the concept [78]. A longer-term NToO operation is envisaged for IceCube and MAGIC and, in case of a discovery, may also become increasingly interesting for other gamma telescopes.

Following another concept, optical follow-up observations of neutrino doublets in IceCube are being prepared [79]. This method can enhance the achievable sensitivity of IceCube to Supernovae and GRB by a factor 2-3. The program is being realized with a small network of automated 1-2 meter telescopes. See also [91, 92].

Two recent papers discuss coincidences between gravitational wave interferometers and neutrino telescopes [80, 81]. Sources of gravitational waves and high energy neutrinos both involve compact objects and matter moving with relativistic speed. Both messengers interact weakly and allow peering into the heart of the regions powering these emissions. A coincidence would be spectacular, with estimated false coincidence rates for IceCube/LIGO being only one event in 435 years [80].

In summary, there is an extremely broad activity on multi-messenger methods, with the potential to increase discovery chances and, in case of a clear discovery, to deepen the understanding of the source.

# OUTLOOK AND CONCLUSIONS

The next years will be key years for opening the neutrino window to the high energy universe. In 2008, remarkable steps in developing high energy neutrino telescopes have been made. IceCube has reached half a cubic kilometer instrumented volume. The recent commissioning of ANTARES defines a milestone towards a next-generation telescope in the Mediterranean Sea, KM3NeT. As emphasized in the recent recommendations of the ApPEC roadmap committee [82], "resources for a Mediterranean detector should be pooled into a single optimized design for a large research infrastructure. The KM3NeT Technical Design Report is expected in late 2009, defining the configuration, deciding between competing technological solutions and providing site arguments. The sensitivity of KM3NeT must substantially exceed that of all existing neutrino detectors including IceCube." [83]. The committee assumes start of construction in 2012 and a five-year installation period.

One may ask the question what "substantially" means. The answer will likely depend on early discoveries (or non-discoveries) with IceCube. In the following I give my own assessment. IceCube is presently entering a region with realistic discovery potential, but theoretical estimates suggest that it may just scratch the regions of discovery. If the integrated extragalactic flux is lower than existing upper "diffuse" limits, the sensitivity of cubic kilometer telescopes is likely insufficient to detect more than a handful of extragalactic sources [84]. Estimates for galactic sources suggest a similarly tantalizing situation. If IceCube would not have identified first sources in, say, 2010, a Mediterranean detector should envisage at least a factor five improvement in sensitivity for $E^{-2}$ sources. Given the budget constraints, this can be reached only at the expense of low-energy sensitivity, i.e. covering a larger volume by using larger spacing between photo-sensors. (A large-spacing array is the approach of the planned GVD in Lake Baikal, although also only with a volume not larger than a cubic kilometer). Figure 11 (taken from [85]) shows detection rates for neutrino sources with a flux $F(> 1\text{ TeV}) = 10^{-11}$ ν cm$^{-2}$ s$^{-1}$ in a km$^3$ detector. This flux corresponds to the cumulative 5σ sensitivity estimated for IceCube in early 2010. It is obvious that the optimal energy range is at 10 TeV to 1 PeV, where the flux exceeds the atmospheric background. Sources with low-energy cut-off energies could be only detected if their flux normalization is higher. For this purpose a low-energy sub-array (like DeepCore for IceCube) is sufficient; instrumentation of a full kilometer with a spacing tailored to energies smaller than 1 TeV might be just waste of resources.

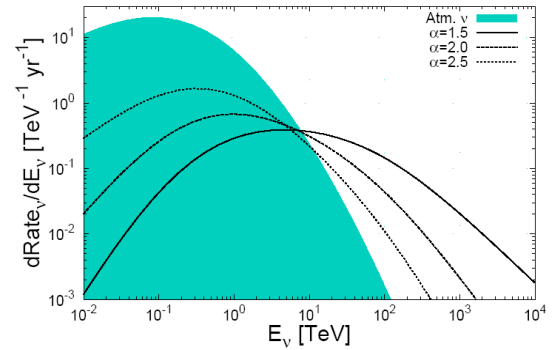

**Fig.11:** Differential detection rates for neutrino sources with flux $F(> 1\text{ TeV}) = 10^{-11}$ ν cm$^{-2}$ s$^{-1}$ in a km$^3$ detector, assuming 3 different spectral indices [85]. Courtesy A. Taylor.

In this review, alternative methods for energies above 100 PeV have not been discussed – see [86] for an experimental review and [93] for signal expectations. Here, new territory has been entered by the South Pole radio detector RICE [87] and by Auger [88] (see also Fig.8) and is also expected to be explored by the Antarctic balloon radio experiment ANITA [89]. A signal indication from one of these experiments would strengthen the motivation to build large dedicated arrays using acoustic and/or radio techniques deep underwater or ice [90].

# ACKNOWLEDGMENTS

I thank the organizers for inviting me to the Gamma-2008 conference in Heidelberg. I am indebted to my colleagues from the IceCube collaboration for cooperation and many useful discussions and A. Kappes, U. Katz, M. Kowalski and P. Lipari for helpful comments on the manuscript.